\documentclass[twocolumn,english,aps,floatfix,superscriptaddress,amsfonts,amssymb,superscriptaddress,prb]{revtex4-1} 
\usepackage{color}
\usepackage[switch,columnwise]{lineno}
\usepackage{graphicx}
\usepackage{amsmath}
\usepackage{latexsym}
\usepackage{epstopdf}
\usepackage{amsmath,amssymb}
\usepackage{algorithm}
\usepackage{algpseudocode}

\usepackage{multirow}
\usepackage{mathtools}
\usepackage[dvipsnames]{xcolor}
\usepackage[english]{babel}
\usepackage{amsthm}
\usepackage{physics}
\usepackage[normalem]{ulem} 

\newcommand{\beq}{\begin{equation}}
\newcommand{\eeq}{\end{equation}}

\def\bra#1{\langle #1|}
\def\ket#1{|#1\rangle}
\DeclareMathOperator*{\argmax}{argmax}

\newcommand{\z}{\mathbf{z}}
\newcommand{\h}{\mathbf{h}}
\newcommand{\vis}{\mathbf{v}}
\newcommand{\bTheta}{\mathbf{\Theta}}

\newtheorem{theorem}{Theorem}

\newtheorem{proposition}[theorem]{Proposition}

\usepackage{tikz}
\usepackage{xcolor}
\usetikzlibrary{patterns}
\allowdisplaybreaks

\usepackage{comment}

\newcommand{\WRP}{\par\qquad\enspace}

\begin{document}

\newcommand\sona[1]{{\color{purple}\textbf{Sona: #1}}}

\title{Many-body localized hidden generative models}
\author{Weishun Zhong}
\email{wszhong@ias.edu Present address: School of Natural Sciences, Institute for Advanced Study, Princeton, NJ, 08540, USA}
\affiliation{Department of Physics, Massachusetts Institute of Technology, Cambridge, MA, 02139, USA}
\affiliation{IBM Quantum, MIT-IBM Watson AI Lab, Cambridge MA, 02142, USA}
\color{black}

\author{Xun Gao}
\affiliation{Department of Physics, Harvard University, Cambridge, Massachusetts 02138, USA}

\author{Susanne F. Yelin}
\affiliation{Department of Physics, Harvard University, Cambridge, Massachusetts 02138, USA}

\author{Khadijeh Najafi}
\affiliation{IBM Quantum, IBM T.J. Watson Research Center, Yorktown Heights, NY 10598 USA}

\begin{abstract}


Born machines are quantum-inspired generative models that leverage the probabilistic nature of quantum states. Here, we present a new architecture called many-body localized (MBL) hidden Born machine that utilizes both MBL dynamics and hidden units as learning resources. We show that the hidden units act as an effective thermal bath that enhances the trainability of the system, while the MBL dynamics stabilize the training trajectories. We numerically demonstrate that the MBL hidden Born machine is capable of learning a variety of tasks, including a toy version of MNIST handwritten digits, quantum data obtained from quantum many-body states, and non-local parity data. Our architecture and algorithm provide novel strategies of utilizing quantum many-body systems as learning resources, and reveal a powerful connection between disorder, interaction, and learning in quantum many-body systems.


\end{abstract}

\maketitle
%
\section{Introduction}

The computational power of quantum processors is the subject of considerable amount of recent research, in particular with regard to scaling and a potential quantum advantage \cite{IBM_eagle,arute2019quantum,IBM2019,Zhong2020,Gong2021,wu2021strong,ebadi2021quantum}. While the advent of a fully error corrected quantum computer requires yet another milestone, the immediate application of noisy quantum hardware with a clear advantage over classical computation becomes even more crucial. In this regard, the interface of quantum computing and machine learning has been increasingly brought into focus. 
For example, the rise of hybrid variational algorithms, such as variational quantum eigensolvers (VQE) \cite{VQE} and the quantum approximate optimization algorithm (QAOA) \cite{QAOA}, which use a parametrized quantum circuit as variational ans"atze and optimize the parameters classically, has been considered particularly promising as they aim to obtain heuristic and approximate solutions.

However, the exponential dimension of the Hilbert space and the random characteristics of parametrized quantum circuits makes their training very challenging  due to the existence of barren plateaus\cite{Jarrod_QNN}. More recently, yet another approach to quantum machine learning has emerged, which is known as  brain-inspired\cite{QNC_Grollier,QMemresistor,Gonzalez-Raya1,Gonzalez_Raya2,Torrontegui}. One interesting category consists of quantum reservoir computing (QRC) where a fixed reservoir geometry scrutinizing the unitary dynamics of an interacting quantum system allows versatile machine learning tasks \cite{QNC_Fuji1,QNC_Fuji2,QRNN_Ryd,xia2022reservoir}. While QRCs have shown many advantages, they are mainly appropriate for discriminative tasks such as classification or regression. 

The goal of generative models, however, is to learn an unknown data probability distribution $p_{\text{data}}$ in order to subsequently sample from $p_{\text{data}}$ and thus generate new and previously unseen data. Such tasks can, for example, be performed by the recently introduced Born machines\cite{Born_MPS,Born_TTN}. Born machines for many-body problems have early on shown to be successful in conjunction with tensor network state ans\"atze.  The elements of these matrix product states or tree tensor networks and their bond dimensions can be optimized during training to effectively approximate $p_{\text{data}}$\cite{Born_MPS,Born_TTN,BEBM_Abigail}. While Born machines have also been tested with parameterized quantum circuits \cite{Born_PQC}, we address here the question of whether there are other quantum many-body states that can be used as anasatz for Born machine to any advantage.  

Quantum many-body systems display many phases in the presence of disorder, in particular, the break-down of thermalization and thus localization of the wavefunction in the so-called \emph{many-body localized} (MBL) phase. Here, emergent integrals of motion can be utilized as quantum memories\cite{Huse2013}. The failure of such systems to anneal\cite{altshuler2010anderson} has inspired their use in QRC\cite{xia2022reservoir} for learning tasks, with particular enhancement close to the phase transition\cite{Martinez2021}.

Here, we extend quantum inspired generative models into the MBL phase, and introduce a hidden architecture to increase the representation power of our generative model. While recent work has also studied Born machines in the MBL phase\cite{tangpanitanon2020expressibility}, using a similar quenched approach, our work differs in the hidden architecture and the characterization of learnability and expressibility. 

    \begin{figure*}[ht]
     \centering
            \includegraphics[scale=0.15]{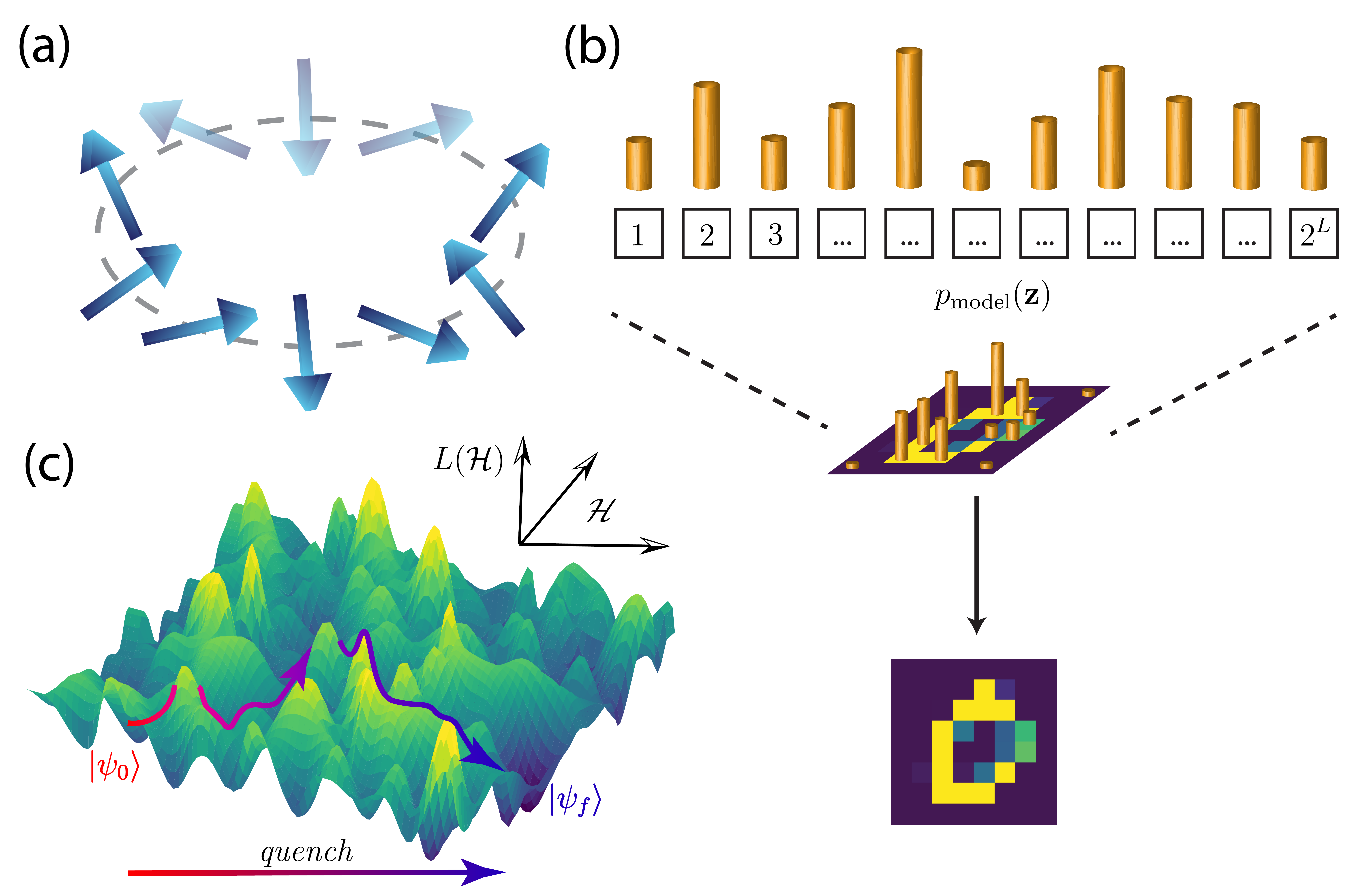}
            \caption{Illustration of the MBL hidden Born machine. (a) XXZ spin chain in 1D with periodic boundary condition. The faded color spins are the hidden units $h_i$, and the solid color spins are the visible units $v_i$. (b) The probability distribution of finding individual states in the z-basis represents the model distribution for the generative model, which are coded as normalized pixel values of an image. (c) An illustration of the loss landscape defined by our hidden MBL Born machine. The training is done by optimizing disorder configurations in the Hamiltonian
            during each quantum quench, which is then used to evolve the initial state $\ket{\psi_0}$ over successive layers of quenches toward a final state $\ket{\psi_f}$ which gives rise to the desired model distribution. }
            \label{fig:model schematics}
        \end{figure*}    

In this article, we introduce the hidden Born machine in section \ref{Hidden-BM}, and describe our training algorithm in section \ref{sec:Training}.
To illustrate the effect of hidden units, we introduce the randomly driven Born machine in section \ref{sec:RDBM} and compare its performance with the hidden Born machine by learning patterns of MNIST hand written digits. Next, in section \ref{sec:different_phases} we investigate the learning power of the hidden Born machine both in the thermal phase and the MBL phase, and numerically show that the thermal phase fails to learn data obtained from quantum systems either in MBL or in thermal phase. Tracking von Neumann entanglement entropy and Hamming distance during training suggests that localization is crucial to learning. In section \ref{sec:parity} we further show that while the hidden Born machine trained in the MBL phase is able to capture the underlying structure of the parity data, a hidden Born machine trained in the Anderson localized phase fails to do so, shedding light on the fact that the interplay between interaction and disorder plays an important role in learning. Finally, we conclude and discuss possible direction for future works. 
\section{Hidden Born machines}
\label{Hidden-BM}
Born machine \cite{liu2018differentiable,cheng2018information,benedetti2019generative,benedetti2019parameterized,coyle2020born} is a generative model that parameterized the probability distribution of observing a given configuration $\z$ of the system according to the probabilistic interpretation of its associated quantum wavefunction $\psi(\z)$,
\begin{equation}
\label{eqn:born}
    p_{\text{Born}}(\z) = \frac{|\psi(\z)|^2}{\mathcal{N}},
\end{equation}
where $\mathcal{N}=\sum_{\z}|\psi(\z)|^2$ is the overall normalization of the wavefunction. Note that $\mathcal{N}$ is only required in tensor network ans\"atze but not in physical systems. Training of the Born machine is done by minimizing the discrepancy between the model distribution $p_{\text{Born}}(\z)$ and the data distribution $q_{\text{data}}(\z)$.


In the language of Boltzmann machine\cite{smolensky1986information,hinton2006reducing,ackley1985learning}, the units that are used for generating configurations are called `visible'. Meanwhile, adding `hidden' units prove to be a powerful architecture for the Boltzmann machine as it provides a way to decouple the complex interaction among the visible units at the expense of introducing interaction between the hidden and the visible units\cite{mehta2019high,gao2017efficient}.
In Eqn.\eqref{eqn:born}, all units of the system are used to generate configurations that are compared against data and therefore all units are visible. In a similar spirit, we introduce hidden units to the Born machine by defining the probability distribution of observing a given visible spin configuration $\z$ to be its expectation value in $z$-basis after tracing out the hidden units, 
\begin{equation}
\label{eqn:hidden}
    p_{\text{hidden}}(\z) = \Tr\rho_{\text{vis}} \Pi_Z,    
\end{equation}
where
\begin{equation}
\label{eqn:rho_vis}
     \rho_{\text{vis}} = \Tr_{h} \ket{\psi}\bra{\psi},
\end{equation}
is the reduced density matrix for the visible units, and $\Pi_Z = \ket{\z}\bra{\z}$ is the projection operator onto the $z$-basis of the visible part of the system (see Fig.\ref{fig:model schematics}(a) for an illustration of our model). Note that normalization is implicit in Eqn.\eqref{eqn:rho_vis} for $\rho_{\text{vis}}$ to be a density matrix. 


The hidden units offer two main advantages:
\begin{itemize}
    \item  (1) when traced out, the hidden units provide an optimization advantage by acting as an effective heat bath for the visible units, in such a way that the systems are less prone to get trapped in local minima;
    \item (2) the hidden Born machine in Eqn.\eqref{eqn:hidden} offers expressive power advantage over the basic Born machine in Eqn.\eqref{eqn:born}.
\end{itemize}

We numerically confirm the mechanism in (1) by comparing the performance between the hidden Born machine and a regular Born machine with an artificially introduced heat bath in Section \ref{sec:RDBM}. For (2), the hidden units provide additional degrees of freedom to parameterize generic probability distributions in an efficient way (see more in Appendix \ref{app:secA}).

\section{Many-body localization phase}
\label{Hidden-BM-expressible}

Previously, different ans\"atze for $\ket{\psi}$ has been introduced for the Born machine, notably tensor networks states and states prepared by both digital quantum circuits and analog quantum many-body systems\cite{Born_MPS,Born_TTN,Born_PQC,tangpanitanon2020expressibility}. In this paper, we will be adopting the latter approach, and focus on a specific type of quantum many-body systems that admits a many-body localization (MBL) phase. 
In the following, we show that the MBL phase has two prominent effects on the training dynamics: 
\begin{itemize}
    \item (1) stabilizing training trajectories: On one hand, the effective heat bath from the hidden units assists the system in escaping local minima. On the other hand, this noise might cause the wavefunction to wander wildly around the Hilbert space, making it difficult for the system to find optimal solutions. Localization prevents such chaotic behavior and enhances the trainability of the model, as illustrated in Fig.\ref{fig:compare models}.
    \item (2) resemblance to associative memory: Our quantum generative model exhibits memory characteristics such as pattern recognition ability, similar to those observed in associative memory architectures inspired by physical systems \cite{hopfield1982neural,krotov2016dense,murugan2015multifarious,zhong2017associative}. We illustrate this in Fig.\ref{fig:learning MNIST}.
\end{itemize}


\subsection{Many-body localized ans\"atze}

It is generally believed that, thermalization in quantum system wipes out the microscopic information associated with the initial state. Even in the case of closed quantum system, the information of initial state quickly spreads throughout the entire system, implying that no local measurements can retrieve those information\cite{Deutsch,srednicki1994chaos}. However, it's known that strong disorder leads to localization, preventing the system to thermalize. Furthermore, the localization manifests itself in the form of memory associated with the lack of transport. While the localization in the presence of strong disorder was first introduced in non-interacting systems by Anderson\cite{Anderson}, more recently, it was shown that the localization and break down of thermalization can also happen in strongly interating systems, leading to new dynamical of phase of matter known as many-body localization (MBL)\cite{MBL1,MBL2}. 

In the MBL phase, eigenstates of the system do not satisfy Eigenstate Thermalization Hypothesis (ETH) and the wavefunctions become localized in the Hilbert space. Such ergodicity breaking renders the system to retain memory of its initial state, and offers advantage in controlling and preparing desired quantum many-body states and has been also realized experimentally\cite{MBL_exp}. The XXZ model of quantum spin chain is well-known to develop a MBL phase when the disorder strength exceeds the MBL mobility edge\cite{MBL_edge}. 

We perform numerical simulation with the XXZ-Hamiltonian
$\hat{\mathcal{H}}_{\text{XXZ}}$ defined as:
    \begin{align}
    \label{eqn:XXZ}
            &\mathcal{\hat{H}}_{\text{XXZ}} = \sum_i^{L-1} J_{xy}(\hat{S}^{x}_{i}\hat{S}^x_{i+1} + \hat{S}^y_{i}\hat{S}^y_{i+1}) + \sum_i ^{L-1} J_{zz} \hat{S}^z_i \hat{S}^z_{i+1},
    \end{align}
where $\hat{S}_i^{\alpha}(\alpha\in \{x,y,z\})$ are Pauli spin 1/2 operators acting on spins $i\in 1,..,L$, and $L=L_v+L_h$ consists of $L_v$ visible units and $L_h$ hidden units. $J_{xy}$ and $J_{zz}$ are couplings in the $xy$ plane and $z$ direction, respectively. Then, we consider a series of $M$ quenches $\hat{\mathcal{H}}_{\text{quench}}(\bTheta_m)$ in the $z$-direction:
    \begin{align}
    \label{eqn:H_total}
        \mathcal{\hat{H}}_{\text{total}}&=\mathcal{\hat{H}}_{\text{XXZ}}+\mathcal{\hat{H}}_{\text{quench}}(\bTheta_m),
    \end{align}
where $\mathcal{\hat{H}}_{\text{quench}}(\bTheta_m) = \sum_i h_i^m \hat{S}^z_i$ and we have denoted the tunable parameters in the system collectively as $\bTheta_m = \{h_i^m\}$. During each quench $m$, $h_i^m$ are drawn i.i.d. from the uniform distribution over the interval $[-h_{d},h_{d}]$, where $h_d$ is the disorder strength. Notice that when $J_{zz}=0$, the model reduces to non-interacting XY model with random transverse field exhibiting single particle localization. Once we turn on the $J_{zz}$ interaction, the spins couple via Heisenberg interaction and MBL phase emerges when $h_c\sim 3.5$ (for $J_{zz}=J_{xy}=1$) \cite{MBL_transition1,MBL_edge,MBL_transition2}. See more details in Appendix \ref{app:MBL_check}.

In section \ref{sec:Training}, we will explain the training algorithm under the time evolution implied by series of quenches in $\hat{\mathcal{H}}_{total}$, and learning through optimizing the values of disordered field $h_i^m$ at each site. 






\section{Training of hidden MBL Born machine}
\label{sec:Training}

\subsection{Learning algorithm}
\label{sec:algorithm}

The basic idea behind the training of Hidden Born machine is the following:
given target distribution $q_{\text{data}}$, and a loss function $\mathcal{L}(p_{\text{model}},q_{\text{data}})$ that measures the discrepancy between model distribution and data distribution, training of the MBL Born machine is achieved through time-evolving the system with the Hamiltonian in Eqn.\eqref{eqn:XXZ}, then optimizing $\bTheta_m$ over $N$ different disorder realizations for each quench $m$. After obtaining the final state at the $M$-th quench, we evaluate the model distribution of the MBL hidden Born machine from Eqn.\eqref{eqn:hidden} and use it as our generative model (see Fig.\ref{fig:model schematics}(c) for an illustration of the learning process). We use Maximum Mean Discrepancy (MMD) loss as our loss function:
\begin{align}
\label{eqn:MMD}
    \mathcal{L}_{MMD} &= \norm{\sum_x p(x)\phi(x)  - \sum_x q(x)\phi(x)}^2,
\end{align}
where $x$ is samples for estimating the MMD loss, and $\phi(x)$ are kernel functions that one can choose (see more details in Appendix \ref{app:MMD}). 

\begin{figure}[h!]
 \centering
        \includegraphics[scale=0.2]{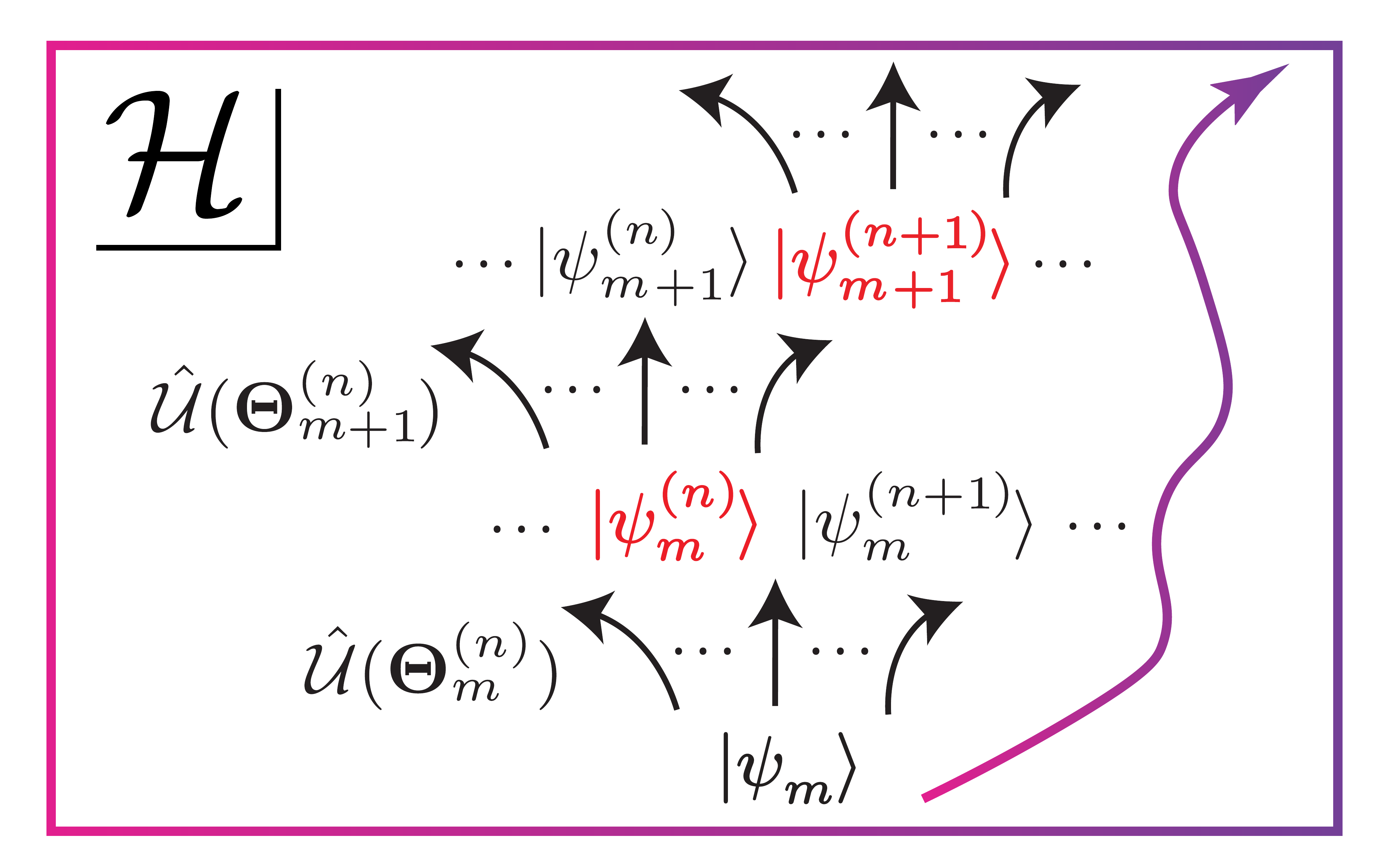}
        \caption{Schematics of the learning algorithm as in Alg.\ref{alg:born}. At the $m$-th quench, we independently evolve $N$ copies of the state $\ket{\psi_{m}}$ with different time-evolution operators $\bTheta_m^{(n)}$ sampled from the same distribution. At the $(m+1)$-th quench, we pick the $\ket{\psi_{m}^{(n)}}$ with the lowest loss value (based on the loss function Eqn.\eqref{eqn:MMD}) from the previous quench as our new starting point and evolve again. As we repeat this process, the learning resembles a directed random walk in the Hilbert space.}
        \label{fig:alg_schem}
\end{figure}

The learning algorithm is summarized by the pseudo-code in Alg.\ref{alg:born} and illustrated in Fig.\ref{fig:alg_schem}.

\begin{algorithm}[H]
\caption{Training of MBL hidden Born machine}\label{alg:born}
\begin{algorithmic}
\State Initialize the system in some initial state $\ket{\psi(\bTheta_{m=0})} \equiv \ket{\psi_0}$ and choose $\bTheta_0 = \mathbf{0}$;
\While{$m < M$}
\While{$n<N$}
\State Sample $\bTheta_{m}^{(n)}$ uniformly from the interval $[-h_{d},h_{d}]$;
\State Time-evolve the state $\ket{\psi_{m+1}^{(n)}} = \hat{\mathcal{U}}(\bTheta_m^{(n)})\ket{\psi_m}$ with \WRP \qquad  $\hat{\mathcal{U}} = \hat{\mathcal{T}}\exp \left(-i\int_0^T dt \hat{\mathcal{H}}_{\text{total}} \right)$;
\State Trace out the hidden units \WRP \qquad $\rho_{m+1}^{(n)} = \Tr_{h} \ket{\psi_{m+1}^{(n)}}\bra{\psi_{m+1}^{(n)}}$;
\State Compute $\mathcal{L}(\bTheta_m^{(n)})$ from \WRP \qquad  $ p_{\text{hidden}}^{(n)}(\z) = \Tr\rho_{m+1}^{(n)}\Pi_Z$;
\State $n \gets n+1$
\EndWhile
\State $\bTheta_{m} = \argmax_{\bTheta_{m}^{(n)}} \mathcal{L}(\bTheta_m^{(n)})$; 
\State $\ket{\psi_{m+1}} = \hat{\mathcal{U}}(\bTheta_m)\ket{\psi_m}$
\State $m \gets m + 1$
\EndWhile
\State Denote the training outcome as $p_{\text{model}}(\z) = \Tr\rho_{M}\Pi_Z$.
\end{algorithmic}
\end{algorithm}



Given the reduced density matrix $\rho_M$ of the $L_v$ visible spins at the final layer of the quench $m=M$, we compute the model distribution from Eqn.\eqref{eqn:hidden}, which gives the probability $p_{\text{model}}(\z)$ of finding each of the $2^{L_v}$ basis states $\z$ in the visible part of the system. For learning image data, we then interpret the probabilities as pixel values (normalized to be within 0 and 1), and reshape it into an image of size $2^{L_v/2} \times 2^{L_v/2}$ (see Fig.\ref{fig:model schematics}(b)). For quantum data, we interpret these probabilities as measurement outcomes obtained from the quantum state. For more details, see Appendix \ref{app:MNIST}.

\subsection{Randomly driven MBL Born machine}
\label{sec:RDBM}
In classical machine learning, stochasticity is found to have the effect of smoothing out loss landscape and helps to avoid local minima \cite{bottou1991stochastic,goodfellow2016deep,mehta2019high}. When introducing the hidden Born machine in Eqn.\eqref{eqn:hidden}, the hidden units are traced out and effectively act as a heat bath for the remaining visible units and provide a source for stochasticity. To understand the source of the learnability advantage provided by the hidden units, we construct a Born machine with an artificially introduced heat bath in this section. We then compare its performance with that of the hidden Born machine, which operates within an effective heat bath induced by the hidden units. \\

Let's consider the Hamiltonian Eqn.\eqref{eqn:XXZ} with applied external random drives $\hat{\mathcal{H}}_{\text{RD}}$ in the $x-$direction (we can also apply random drives in the $xy-$plane and the result will be similar),
\begin{equation}
\label{eqn:randomdrive}
    \hat{\mathcal{H}}_{\text{RD}}(t) = \sum_i d_i^m(t) \hat{S}_i^x.
\end{equation}
To model the heat bath, we would like $\{d_i^m(t)\}$ to be like a white noise,
\begin{equation}
\label{eqn:whitenoise}
    \langle d_i^m (t) d_i^m(0) \rangle = 2D \delta(t),
\end{equation}
where $D$ is the amplitude of the white noise and is proportional to the temperature of the bath. 
In the simulation, we split the driven interval $T$ into intervals of auto-correlation time $\tau$, and require that Eqn.\eqref{eqn:whitenoise} holds for $t>\tau$. Outside of this correlation time, $d_i^m (t)$ is drawn i.i.d. from $\mathcal{N}(0,\sqrt{2D})$. We refer to this model as Randomly Driven Born machine (RDBM). 


\begin{figure}[h!]
 \centering
        \includegraphics[scale=0.17]{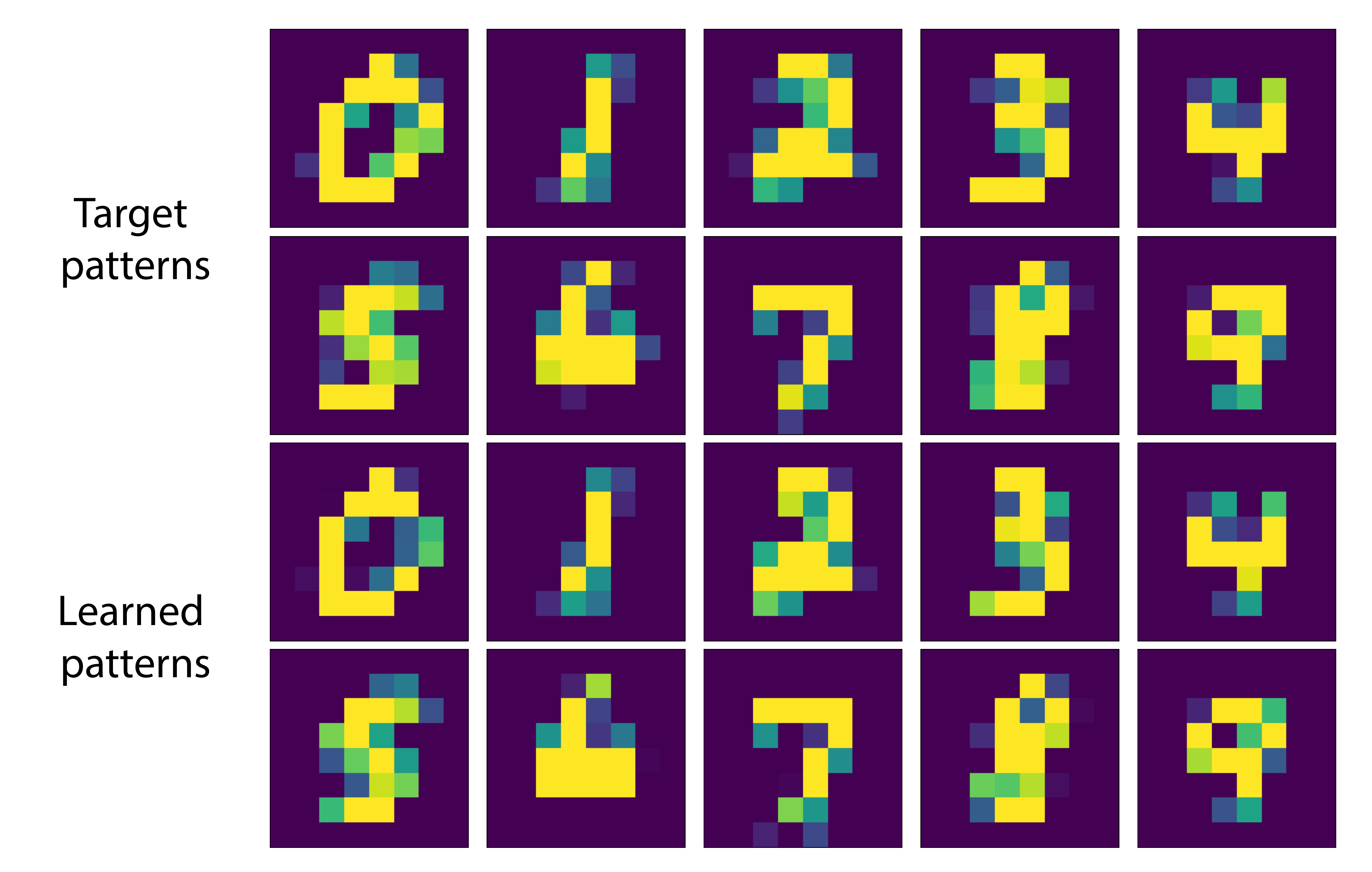}
        \caption{Learning toy MNIST digit patterns. The top two rows are different data instances $q_{\text{data}}$ in our toy MNIST digit patterns dataset. The bottom two rows are the corresponding learning outcome $p_{\text{model}}$ from our MBL hidden Born machine (each digit trained separately).}
        \label{fig:MNIST_0_9}
\end{figure}    


To illuminate on the learning power of the hidden Born machine, here, we compare the three models: the basic Born machine (BM), the RDBM, and the hidden Born machine (hBM). We task all three models with a toy dataset constructed from the images of MNIST dataset\cite{deng2012mnist} (downsampled to $2^{L_v}$ pixels). Our toy dataset consists of mean pixel values across all different styles within a single type of MNIST digit, see `target patterns' in Fig.\ref{fig:MNIST_0_9} (also see Appendix \ref{app:MNIST}).

We perform the training of the hidden Born machine using the algorithm described in Alg.\ref{alg:born}, and show the corresponding learning outcomes in Fig.\ref{fig:MNIST_0_9}. Our results indicate our hidden model is able to learn different patterns of MNIST digits accurately (the result of basic BM and RDBM are omitted). 

\begin{figure}[h!]
\centering
\includegraphics[scale=0.27]{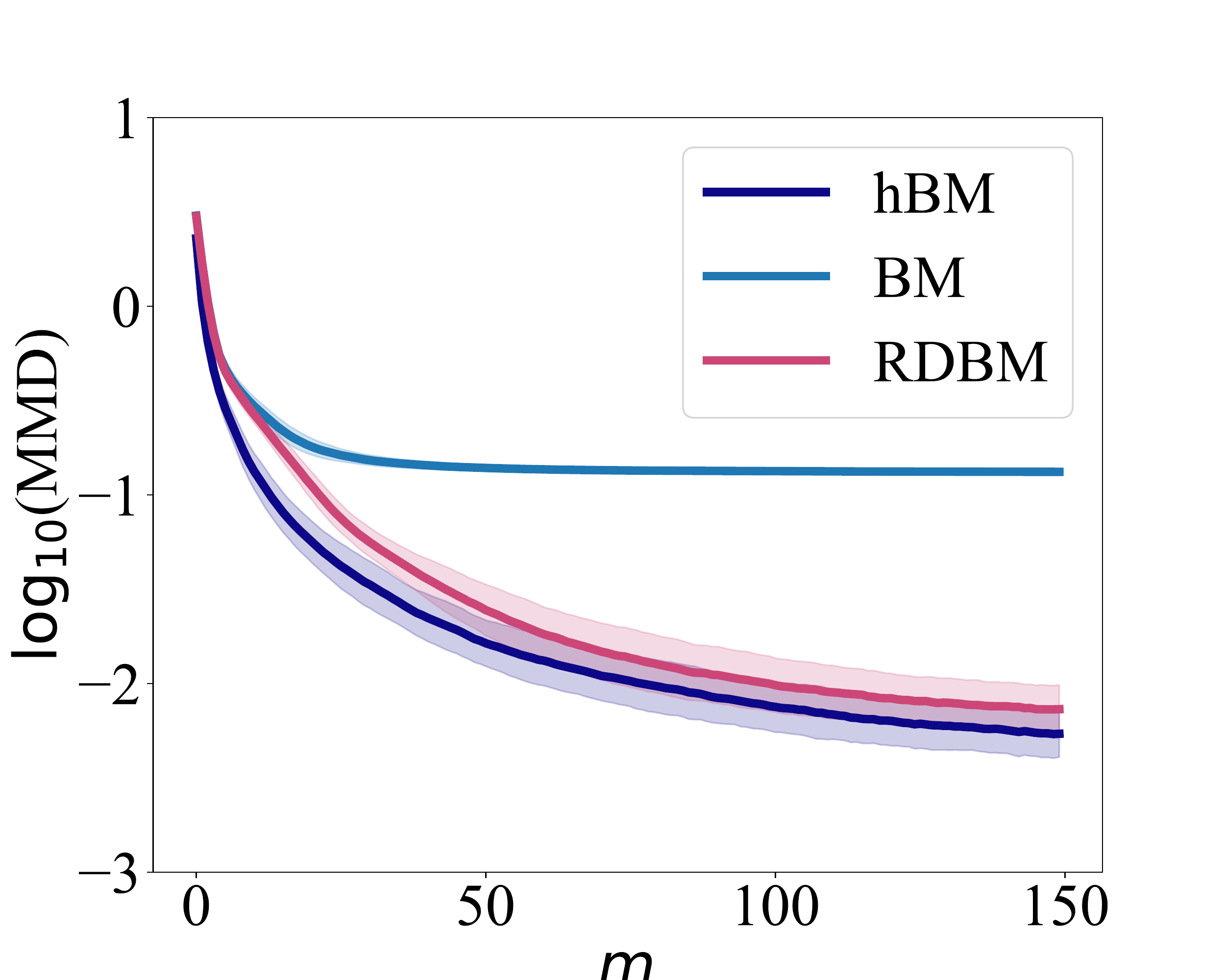}
\caption{Model comparisons. (a) Basic Born machine (BM), Randomly Driven Born machine (RDBM), and hidden Born machine (hBM). Log-MMD loss as a function of quench layer number $m$. The solid curves are averaged over 100 different realizations, with one standard deviation included as the shades. The hidden Born machine achieves the lowest MMD loss throughout and at the end of the training.}
\label{fig:loss_MNIST}
\end{figure}
%
In Fig.\ref{fig:loss_MNIST}, we compare the performance of three models by plotting the MMD loss on the toy MNIST dataset as a function of the number of quenches, $m$. We find that both the RDBM and hBM outperform the BM, with the RDBM and hBM converging to a similar terminal loss value (although the RDBM slightly underperforms the hBM). This observation suggests that the stochasticity introduced by the heat bath indeed improves learning, and the learnability advantage of hidden units is of a thermal nature. However, the RDBM is more computationally intensive. Modeling the white noise as expressed in Eqn.\eqref{eqn:whitenoise} requires discretizing the quench time $T$ into small intervals of $\tau$, which generally increases the total runtime by a factor of $T/\tau$. Therefore, the hBM is more efficient than the RDBM while being more trainable than the basic BM.




\begin{figure*}[ht]
    \centering
    \includegraphics[scale=0.25]{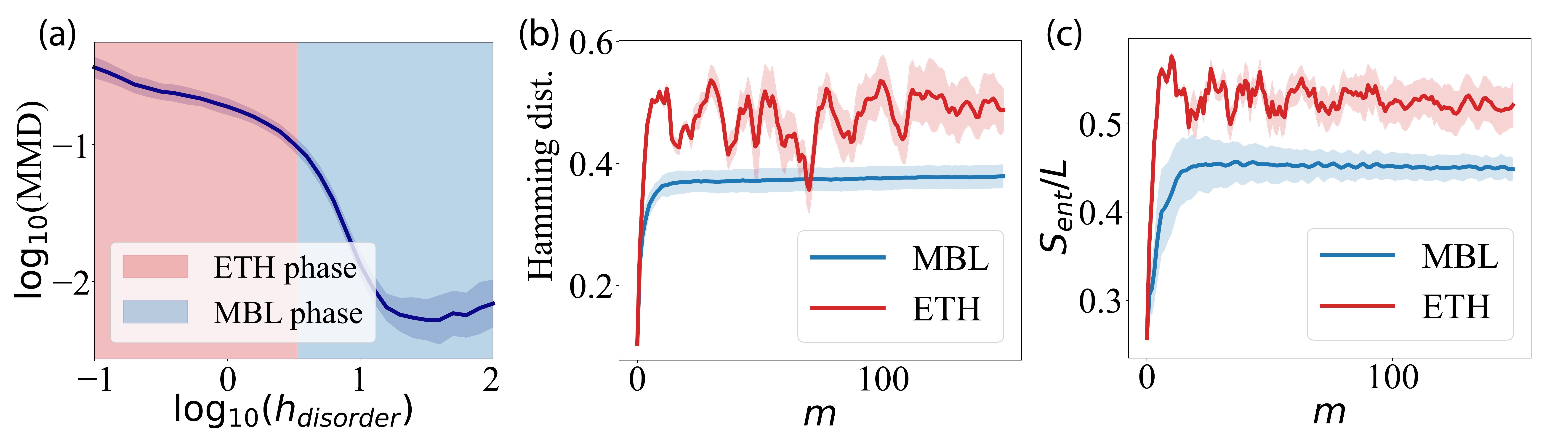} 
    \caption{Training hidden Born machine in thermal and MBL phases. (a) The terminal (at the final layer of quench) MMD loss of hidden Born machine on the toy MNIST task is plotted as a function of disorder strength $h_{d}$. The results are averaged over 100 realizations and one standard deviation is included as shade. (b) Hamming distance with respect to the initial state (normalized by $L$) as a function of quenches $m$. In the thermal phase, states change discontinuously over successive quenches, whereas in the MBL phase states change gradually toward the target state that gives rise to the desired distribution. (c) Entanglement entropy per site as a function of quenches $m$, confirming that our system evolves under dynamics distinctive in the thermal/MBL phases.}
    \label{fig:compare models}
\end{figure*}    

\section{Learnability in different phases}
\label{sec:different_phases}
We have already seen that the hidden Born machine in the MBL phase can properly learn the toy MNIST dataset (Fig.\ref{fig:MNIST_0_9}). An important question arises that whether learning can happen in the thermal phase. In the thermal phase, information spreads throughout the system, which makes it difficult to extract. In the quenched approach as in Eqn.\eqref{eqn:H_total}, the state of the system in the thermal phase changes wildly between successive quenches and effectively only parameters in the last layer of the quench would be trained. In contrast, as the system become more disorderd and enters the MBL phase,  the breakdown of thermalization and emergence of local integrals of motion leads to local memory, which is useful for directing the state toward a target corner of the Hilber space (Fig.\ref{fig:model schematics}(c)). We aim to understand the effect of disorder in learning by comparing the learning ability of the hidden Born machine in the MBL and thermal phases.

In Fig.\ref{fig:compare models}(a), we show a log-log plot of the final quench layer MMD loss on the toy MNIST dataset as a function of disorder strength $h_d$. By varying the disorder strength, the system in Eqn.\eqref{eqn:XXZ} can exhibit both a thermal phase (denoted as ETH) and an MBL phase depending on whether the disorder strength exceeds the critical value $h_c\sim 3.5$ (for $J_{zz}=J_{xy}=1$). We observe that the loss value has a significant change at the transition from the thermal phase (corresponding to $h_d < 3.5$ indicated by pink shade) into the MBL phase (indicated by blue shade). The relatively high value of MMD loss in the thermal phase indicates that the hidden Born machine fails to learn. As we increase the disorder, the MMD loss deep in the MBL phase decreases significantly, indicating better learning power of the MBL phase. We can attribute the better learning power in the MBL phase to the quantum memory and the emergent local integral of motions. In contrary to the thermal phase, the thermalization mechanism wipes out all the information from the initial conditions, as observed similarly in the case of quantum reservoir computing in the MBL phase \cite{xia2022reservoir}.

To better quantify the learning mechanism in the MBL phase, we investigate the time evolution of quantities underlying MBL physics during the quenched steps. First, we investigate the Hamming distance (HD) defined as
\begin{equation}
\label{eqn:Hamming_distance}
\mathcal{D}(t)= \frac{1}{2}-\frac{1}{2L}\sum_i \langle \psi_0|\sigma_i^{z}(t)\sigma_i^{z}(0)|\psi_0\rangle,
\end{equation}
which gives a measure of number of spin flips with respect to the initial state $\psi_0$ normalized by the length of chain $L$. It's expected that in the long time the HD approaches 0.5 in the thermal phase and decreases as one increases the disorder\cite{Hauke_2015}. In Fig.\ref{fig:compare models}(b), we show the trajectory of HD at the end of each quench $\mathcal{D}^m(t=T)$. We observe that, evolving in the thermal phase the HD fluctuates around the value of 0.5 as expected, while in the MBL phase the HD reaches a lower value about 0.33. The more significant fluctuations in the thermal phase indicate that the system retains little information about the most recent quench, and therefore is difficult to be manipulated toward a target state that gives desired probability distribution. In contrast, the relatively small fluctuations in the MBL phase suggest that system changes gradually between successive quenches and is more amenable to directed evolution by quenches.

One hallmark of MBL phase is the logarithmically slow growth of von Neumann entanglement entropy ($S_{\text{ent}}^m=-\rm Tr\rho_m \ln \rho_m$) due to the presence of strong interaction. Notice that $\rho_m$ is the reduced density matrix at quench $m$, obtained by tracing over the complementary part of system with respect to the subsystem of interest. This can be considered as slow dephasing mechanism implying that not all information of initial state survives \cite{MBL_ent1,MBL_ent2,MBL_ent3}. In order to confirm that our system indeed evolves under MBL/thermal dynamics when trained in these two phases, in Fig.\ref{fig:compare models}(c), we track the value of $S_{\text{ent}}^m$ over different quenches. In the MBL phase, $S_{\text{ent}}^m$ shows a quick saturation, while in the thermal phase the entanglement entropy changes significantly from successive quenches, a behavior expected from the thermal phase.

\subsection{Pattern recognition}
\label{sec:pattern recognition}

\begin{figure}[h!]
\centering
\includegraphics[scale=0.25]{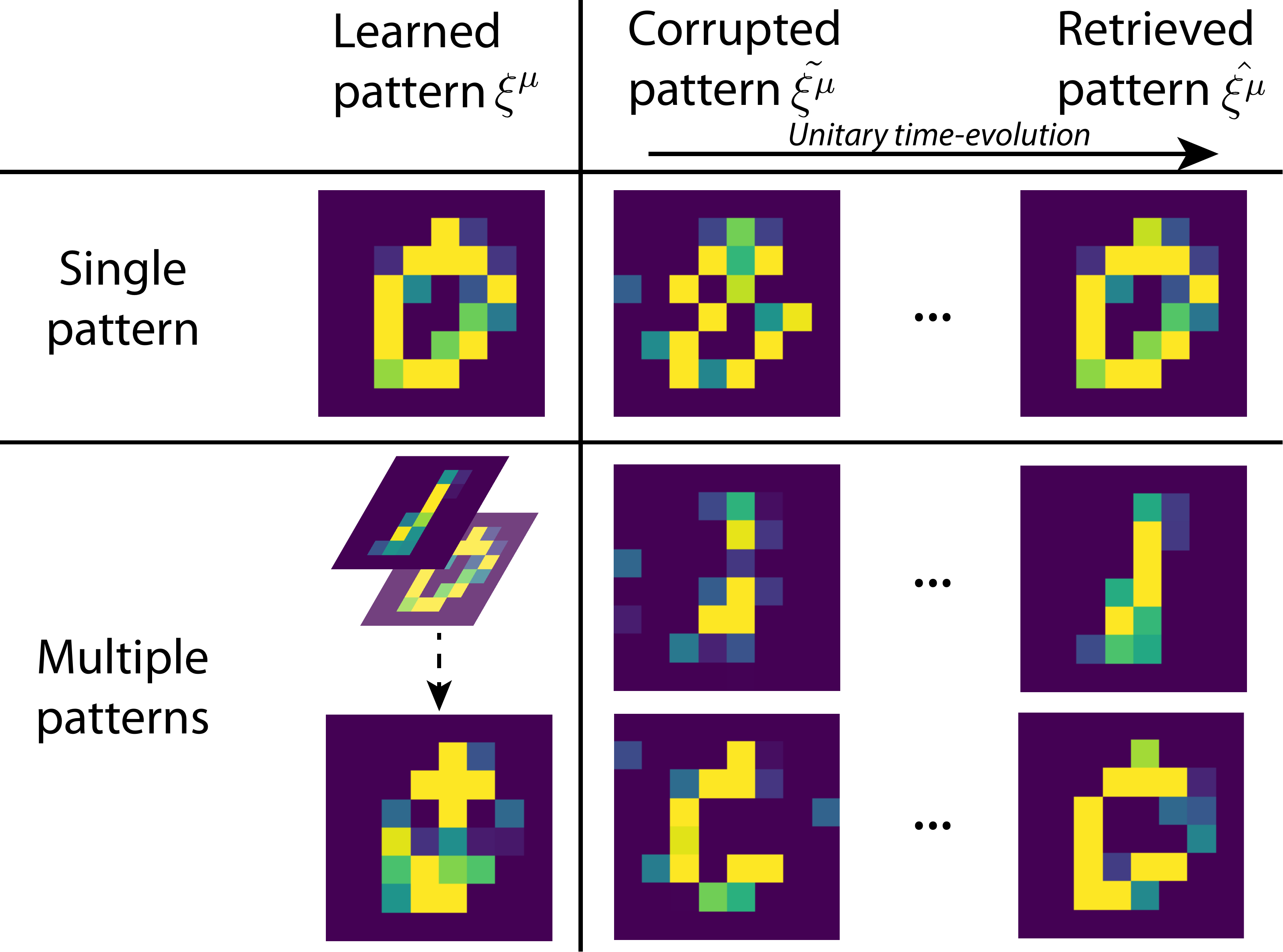}
\caption{Pattern recognition task by the MBL hidden Born machine. Given a corrupted pattern $\Tilde{\xi}^{\mu}$ and its corresponding corrupted state $\ket{\Tilde{\psi}^{\mu}}$, we find the quench layer number $m^*$ such that the intermediate model distribution $p_m^*$ resembles the corrupted pattern most. Then we time-evolve $\ket{\Tilde{\psi}^{\mu}}$ with the series of learned unitaries $\hat{\mathcal{U}}_i$ starting from $m^*$ to obtain the retrieved state $\ket{\hat{\psi}^{\mu}}$, from which we can then obtain the retrieved pattern $\hat{\xi}^{\mu}$. Top row: after learning a single pattern (digit `$0$'), a complete `$0$' can be retrieved from a partially corrupted `$0$'. Bottom row: after learning multiple patterns (superposition of digit `$0$' and `$1$'), complete `$0$' or `$1$' can be selectively retrieved from partially corrupted `$0$' and `$1$', respectively.}
\label{fig:learning MNIST}
\end{figure}

Pattern recognition has been implemented in a variety of analog classical systems ranging from molecular self-assembly to elastic networks \cite{winfree1998algorithmic,qian2010efficient,murugan2015multifarious,zhong2017associative,o2019temporal,stern2020continual,zhong2021machine,stern2022learning}. It is interesting to ask whether quantum systems possesses similar power. In this section, we demonstrate the pattern recognition ability of the MBL hidden Born machine. Here, we take the same toy dataset of MNIST digit patterns as in Fig.\ref{fig:MNIST_0_9}. Each pattern $\xi^{\mu} \in [0,1]^{2^{L_v}}$ is a (normalized) vector in the pixel space, where $L_v$ is the length of the visible units, and $\mu=1,2,..,P$ denotes the pattern index. We encode the patterns into the hidden Born machine by setting $p_{\text{data}} = \sum_{\mu}\xi^{\mu}$ \footnote{While there exists other more sophisticated encoding schemes, here we choose the simplest one for illustration.}. 
Again, we perform the training of the hidden Born machine using the algorithm in Alg.\ref{alg:born} (see first column of Fig.\ref{fig:learning MNIST} for the learned patterns from $p_{\text{model}}$). After training, we obtain the target final state $\ket{\psi_M}$, along with a series of unitaries $\{ \hat{\mathcal{U}}(\bTheta_m) \equiv \hat{\mathcal{U}}_m \}_{m=0}^{M}$ that defines the entire history of intermediate states during successive quenches, $\ket{\psi_m} = \prod_{i=0}^m \hat{\mathcal{U}}_{m-i} \ket{\psi_0}$, which upon tracing out hidden units becomes intermediate model distributions, $p_m = \Tr \Tr_{h} \ket{\psi_{m}}\bra{\psi_{m}}\Pi_Z$. Now given a partially corrupted pattern $\Tilde{\xi}^{\mu}$ and the state $\ket{\Tilde{\psi}^{\mu}}$ that gives rise to this corrupted pattern, $|\Tilde{\psi}^{\mu}(\boldsymbol{z})|^2/\mathcal{N} = \Tilde{\xi}^{\mu}$ (see second column of Fig.\ref{fig:learning MNIST} for examples of corrupted patterns), we can identify the `closest' intermediate state $\ket{\psi_{m^*}}$ where $m^*=\argmax_m \text{MMD}(\Tilde{\xi}^{\mu},p_{m})$. Then we apply unitary time-evolution to the corrupted state $\ket{\Tilde{\psi}^{\mu}}$ using the series of learned unitaries starting from $m^*$ and obtain the `retrieved' state $\ket{\hat{\psi}^{\mu}} \equiv \prod_{i=0}^{M-m^*}\hat{\mathcal{U}}_{M-i} \ket{\Tilde{\psi}^{\mu}}$. We can then compute the corresponding retrieved pattern as $\hat{\xi}^{\mu} \equiv \Tr \Tr_{h} \ket{\hat{\psi}^{\mu}}\bra{\hat{\psi}^{\mu}}\Pi_z$ (see last column of Fig.\ref{fig:learning MNIST} for the retrieved patterns).

\label{sec:MNIST}

As shown in the top row of Fig.\ref{fig:learning MNIST}, in the case of a single pattern (a digit `0'), the MBL hidden Born machine is able to retrieve a complete pattern from a corrupted pattern (a partially corrupted digit `0'). As shown in the bottom row of Fig.\ref{fig:learning MNIST}, in the case of multiple patterns (a superposition of `0' and `1'), the MBL hidden Born machine is able to selectively retrieve complete patterns (`0' or `1') based on the input corrupted pattern \footnote{However, one should note that just like in classical pattern recognition \cite{hopfield1982neural}, if the input pattern gets too corrupted and does not resemble any of the encoded patterns, this procedure will fail.}.

\subsection{Learning quantum dataset}
\label{sec:quantum_dataset}


\begin{figure*}[ht]
    \centering
    \includegraphics[scale=0.2]{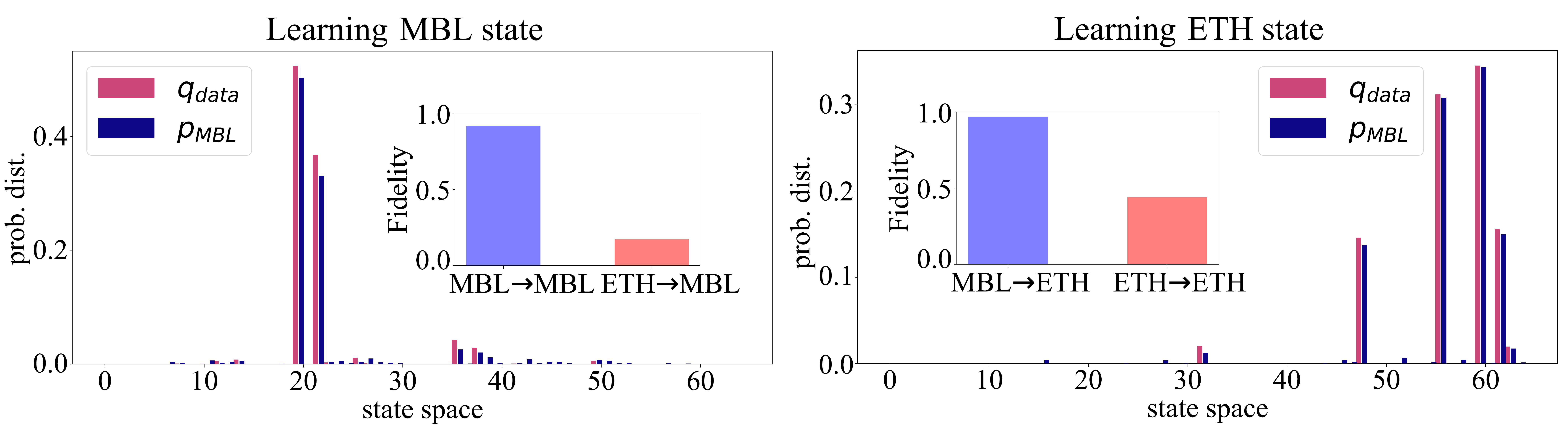} 
    \caption{Learning quantum dataset. Left/right: MBL hidden Born machine trained in MBL phase learns the probability distribution corresponding to an MBL/thermal (denoted as ETH) target state. Insets: classical fidelities between the model and the data distributions. Model trained in the MBL phase has better learning capability than model trained in the thermal phase. }
    \label{fig:learning_quantum_states}
\end{figure*}    


We have demonstrated the power of MBL Born machine in learning classical data of the toy MNIST digit patterns, now we explore the ability of the MBL Born machine in learning data obtained from measurements of quantum states. While quantum state tomography is the standard method for state reconstruction, it becomes a daunting task as the system size increases. In this respect, quantum machine learning has shown great success in learning quantum states from limited amount of data\cite{carrasquilla_reconstructing_2019,torlai_neural_network_2018,tomography_Wang_2020,huang_provably_2021,huang_predicting_2020,randomized_toolbox,Abigail_Enhanced_Born}. In this section, we use the hidden Born machine to learn data obtained from quantum many-body states prepared by Eqn.\eqref{eqn:H_total} subject to a single layer of quench, but with disorder strengths $h_d$ different from the phases that the hidden Born machine is trained in.


In Fig.\ref{fig:learning_quantum_states}, we demonstrate the learning ability of Born machine in the thermal and MBL phase. In Fig.\ref{fig:learning_quantum_states} left/right, we compare the measurement outcome sampled from the exact simulation $q_{\text{data}}$ in MBL/thermal phase (denoted as ETH)  (shown in purple), with those learned via hidden Born machine trained in MBL phase (shown in blue). In the insets we show the classical fidelity between the model distribution $p$ and data distribution $q$, $F(p,q)=\left( \sum_i \sqrt{p_i q_i} \right)^2$. We see that the hidden Born machine trained in MBL phase is able to capture the underlying probability distribution obtained from both the MBL and thermal phases with high fidelity ($\sim 0.98$), while the hidden Born machine trained in thermal phase fails to learn either. Notice that in order to learn the quantum state, one needs to perform measurement in the informationally-complete basis as reported in Ref.\cite{Abigail_Enhanced_Born}.

\subsection{Learning parity dataset}
\label{sec:parity}

\begin{figure}[h!]
\centering
\includegraphics[scale=0.3]{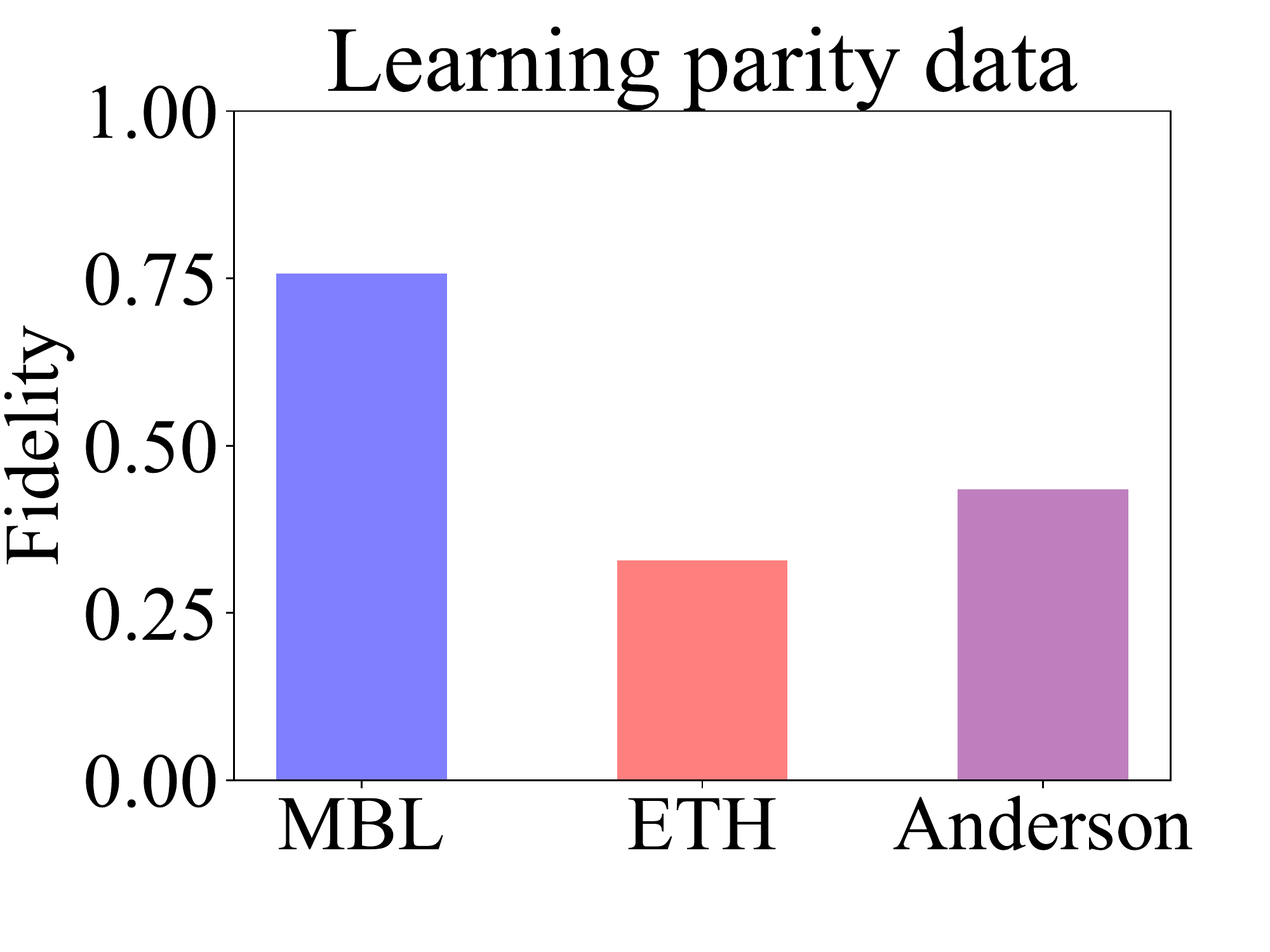}
\caption{Learning parity dataset. Different bars in the horizontal-axis correspond to model trained in the MBL, thermal, and Anderson localized phases, respectively. Vertical-axis shows the classical fidelity of the model. Model trained in the MBL phase exhibits the highest fidelity despite the dataset is highly nonlocal. Comparing model performances in three phases suggest that both disorder and interaction are important for learning.}
\label{fig:learning parity}
\end{figure}

In the previous sections, we have discuss the role of localization and emergent memory in learning various datasets, however, the role played by interaction in the many-body localized phase remains unclear. To shed light into the role of interaction and its interplay with disorder, here, we investigate the power of MBL phase in learning parity dataset and compare it with both thermal and Anderson localized phase which can be obtained by setting $J_{zz}=0$ in Eqn.\eqref{eqn:XXZ}. Here, we consider the even parity dataset, which is defined as set of bit-string $(b_1,b_2,..,b_L)$ of length $L$ with $b_i\in\{0,1\}$, such that the parity function $\Pi(b_1,b_2,..,b_L):=\sum_{i=1}^{N} b_i \mod 2 $ is equal to $0$. While this is a classical bitstring, it appears as measurement outcome of particular quantum observables in certain basis such as measurement outcome of GHZ state in the $x-$basis.

Previous studies has indicate challenging learning on this dataset, in particular training the Born machine based on MPS with gradient descent optimization schemes has encountered failures \cite{Najafi_GHZ_nonlocal}, while quantum inspired optimization schemes such as density matrix learning has shown great success with the caveat in their scaling\cite{Bradley_2020}. Here, we investigate the power of our hidden Born machine across various phases in learning the parity dataset. Our numerical results (Fig.\ref{fig:learning parity}) demonstrates the interesting fact that both the MBL phase and Anderson localized phase show better performance compare to the thermal phase. The better learning performance in these two phases suggest that the emergence of integral of motion and memory plays an important role in learning. We further notice that the MBL phase has a better performance even though the Anderson localized phase is known to have better memory. In the latter the strong localization prevents the transport of information across the system, leading to a lesser learning power. While the value of fidelity around $F_{\text{MBL}}=0.75$ is not too high, reflecting the hardness of learning the parity dataset, our MBL hidden Born machine still shows a better performance compare to MPS Born machine which was reported a fidelity of $F_{\text{MPS}}=0.48$\cite{Najafi_GHZ_nonlocal}. Our numerical results indicate that both disorder and interaction are crucial for the successful learning of our model. The MBL phase enhances the learnability of the hidden Born Machine because it incorporates these two properties. However, given that the XXZ model is not the only one with such characteristics, we anticipate that the hidden Born Machine, when evolved under other Hamiltonians with disorder and interaction - such as spin-glass Hamiltonians \cite{kjall2014many} - would exhibit similar learnability. Intuitively, one can think of the Hidden Born Machine as carving a path through the Hilbert space, capitalizing on both local memory and interaction to reach the target state.

\section{Conclusion and outlook}
In this work, we have introduced the hidden MBL Born machine as a powerful quantum inspired generative model. Although parameterized quantum circuit has become one of the focal point in the realm of quantum machine learning, their training scheme poses many challenges as one requires to search in an exponential Hilbert space, which resembles finding a needle in haystack\cite{Jarrod_QNN}. While other variational algorithms such as QAOA offer a different scheme of finding solution in Hilbert space which is akin to adiabatic computing, here, by utilizing unique properties of MBL phase such as localization and memory, we develop a Born machine evolving under MBL dynamics such that by optimizing over values of disorder at each site we can reach a desired target state in the Hilbert space. 


In this work, we aimed to answer two key questions, namely, whether MBL phase and hidden units can be used as resource for learning, and what is the underlying mechanism of learning. By performing various numerical experiments in different phases, with and without hidden units, we show that successful learning relies on three key factors, hidden units, disorder, and interaction. 

Our work opens up a new horizon in utilizing exotic quantum phases of matter as quantum inspired generative models. While we have explored the role of disorder in the MBL phase, an immediate question that follows is whether other disordered quantum phase would be capable of learning, which is left for future work. In addition, although we show that our model contains states that cannot be simulated in polynomial time by classical computer, it remains an important question of designing efficient quantum algorithms to prepare such states. Similarly, it remains an open question whether the MBL hidden Born machine can learn other classically intractable distributions \cite{fefferman2017exact,coyle2020born,movassagh2023hardness}, such as the distribution generated in quantum supremacy experiments \cite{arute2019quantum,IBM2019,Zhong2020,Gong2021,wu2021strong}. Furthermore, our quenched Born machine resembles specific adiabatic schedule, and whether we can utilize our model as quantum variational algorithm awaits further investigation. Although we have quantified the learning mechanism during the training by tracking both local and non-local quantities such as Hamming distance and entanglement entropy, more quantitative studies such as the existence of Barren Plateau and over-paramtrization in the context of quantum kernel learning remains an important question for future study\cite{Jarrod_QNN,Taylor_barren,liu2022analytic}.

\begin{acknowledgements}

The authors would like to thank Nishad Maskara and Nicole Yunger Halpern for useful discussions. W.Z. acknowledges supports from IBM Research (Quantum) summer internship program. SFY would like to acknowledge funding from NSF (through CUA and Q-SeNSE) and AFOSR.

\end{acknowledgements}




\newpage

\bibliography{ref.bib}

\appendix

\section{Expressibility of MBL hidden Born Machine}
\label{app:secA}

\subsection{Expressibility of hidden units}
The hidden Born machine Eqn.\eqref{eqn:hidden} generalizes the basic Born machine (BM) defined by Eqn.\eqref{eqn:born}, in the sense that the class of probability distributions expressible by the basic Born machine is a subset of that of the hidden Born machine. 

\begin{proposition}
\label{prop_1}
For the same set of visible spins $\vis$, let $p_{\text{BM}}(\z)$ denote the model distribution realized by the basic Born machine, and $p_{\text{hBM}}(\z)$ denote the model distribution realized by the hidden Born machine, then $\{p_{\text{BM}}(\z)\}\subseteq \{p_{\text{hBM}}(\z)\}$.

\end{proposition}

Prop.\ref{prop_1} further suggests that the minimum achievable training loss for the hidden Born machine is less than or equal to that of the basic Born machine, a property that we will confirm numerically in section \ref{sec:Training}. 

\label{app:proof_prop1}
In the following, we assume only that the visible and hidden part couple through an interaction term in the Hamiltonian.

Let's consider a basic Born machine consisting of only visible units $\vis=\{v_i\}$, with Hamiltonian $\hat{\mathcal{H}}_v$. Now consider adding hidden units $\h=\{h_j\}$ to the system with Hamiltonian $\hat{\mathcal{H}}_h$ and the hidden units couple with the visible ones through an interaction Hamiltonian $\hat{\mathcal{H}}_{\text{int}}$. The full Hamiltonian can be written as 
\begin{equation}
\label{eqn:fullH}
    \hat{\mathcal{H}}_{vh}[\vis,\h] = \hat{\mathcal{H}}_v[\vis] + \hat{\mathcal{H}}_h[\h] + \hat{\mathcal{H}}_{\text{int}}[\vis,\h],
\end{equation}
where all the $\hat{\mathcal{H}}$'s in general can be time-dependent. Let's assume that the basic Born machine model is described by just the visible part of Hamiltonian in Eqn.\eqref{eqn:fullH}, $\hat{\mathcal{H}}_{\text{BM}} = \hat{\mathcal{H}}_v(\bTheta^{\text{BM}})$, and the hidden Born machine is described by the full Hamiltonian, $\hat{\mathcal{H}}_{\text{hBM}} = \hat{\mathcal{H}}_{vh}(\bTheta^{\text{hBM}})$, where $\bTheta^{\text{BM}}$ and $\bTheta^{\text{hBM}}$ denotes the parameters in the Hamiltonian to be optimized during learning.

\begin{proof}
Let's denote the initial state for the BM as $\ket{\psi_{0}^v}\in \mathcal{H}_{v}$. Let $\hat{\mathcal{U}}_{v} = \hat{\mathcal{T}}\exp \left(-i\int_0^T dt \hat{\mathcal{H}}_{v} \right)$. Then, the final state of BM is $\ket{\psi_f^v} = \hat{\mathcal{U}}_{v}\ket{\psi_0^v}$.
Choose an initial product state for the hBM, $\ket{\psi^{vh}_0} = \ket{\psi_0^v} \otimes \ket{\psi_0^h} \in \mathcal{H}_{v} \otimes \mathcal{H}_{h}$ for some $\ket{\psi_0^h} \in \mathcal{H}_h$. Choose $\bTheta^{\text{hBM}}$ to be such that $\hat{\mathcal{H}}^{\text{hBM}}_{v} = \hat{\mathcal{H}}^{\text{BM}}_{v}$, and $\lvert\lvert \hat{\mathcal{H}}^{\text{hBM}}_{v} \rvert\rvert \gg \lvert\lvert \hat{\mathcal{H}}^{\text{hBM}}_{\text{int}} \rvert\rvert$.

Then, we have 
\begin{align}
    \ket{\psi_f^{vh}} \approx \hat{\mathcal{U}}_{v} \ket{\psi_0^v} \otimes \hat{\mathcal{U}}_h \ket{\psi_0^h} = \ket{\psi_f^v} \otimes \ket{\psi_f^h}
\end{align}
where we have defined $\ket{\psi_f^h} \equiv \hat{\mathcal{U}}_h \ket{\psi_0^h}$. With this choice, now $\rho_{\text{vis}} = \Tr_h \ket{\psi_f^{vh}} = \ket{\psi_f^v}\bra{\psi_f^v}$, and $p_{\text{hBM}}(\z) = \Tr \rho_{\text{vis}}\Pi_Z = |\psi_f^v(\z)|^2 = p_{\text{BM}}(\z)$, where $p_{\text{BM}}$ is automatically normalized ($\mathcal{N}=1$) for physical systems as in our case. Therefore, the class of probability distributions described by BM is contained in hBM.

\end{proof}

\subsection{Mapping XXZ chain into Ising model}
\label{app:proof_prop2}

There has been extensive studies on the expressive power of quantum models. In particular, quantum computational advantage for sampling problem has been proved (based on standard computational complexity assumptions) in a translation-invariant Ising model \cite{Gao_2017}. In particular, we show that the XXZ model in 2-dimension, with proper choice of disorder parameters, can be reduced to an Ising model that contains brickwork state that is classically intractable Ref.~\cite{Gao_2017}. 

\begin{proposition}
The XXZ model in 2D subject to quench in z-direction can be reduced to an Ising model.
\end{proposition}

We have recovered the case in Ref.~\cite{Gao_2017}. Note that for the proof in Ref.~\cite{Gao_2017} to work, one also need to initialize the system in all $\ket{+}$ states and subsequently perform all measurements in the $x-$basis. While our numeric are mostly restricted to the 1-dimension case as it can be studied by exact diagonalization, the XXZ model can be realized in any dimensions. This classically-hard instance implies that our model cannot be simulated in polynomial time by a classical computer and therefore offers an advantage in its expressive power. 

\begin{proof}

In 2D, 

\begin{equation}
    \mathcal{\hat{H}}_{\text{XXZ}} = \sum_{\langle i,j \rangle} J_{xy}(\hat{S}^{x}_{i}\hat{S}^x_j + \hat{S}^y_{i}\hat{S}^y_j) + \sum_{\langle i,j \rangle} J_{zz} \hat{S}^z_i \hat{S}^z_{j},
\end{equation}
where the interactions are between nearest neighbours. During a quench $\hat{\mathcal{H}}_{\text{quench}}$ of duration $t_m$, we can divide the disorder into a time-dependent and a time-independent part, 
\begin{equation}
    h_i^m(t) = J_i^m(t) + B_i^m.
\end{equation}

In the case of bipartite lattice, we can partition the vertices into two partitions, and denote the sites in one partition as $\mathcal{K} = \{k_1, k_2,..,k_{L/2}\}$ and another partition as $\mathcal{N} = \{n_1,n_2,...,n_{N/2}\}$ (assuming $N$ even). For example, in the case of a square lattice, $\mathcal{K}$ and $\mathcal{N}$ correspond to the blue and white sites of the checkerboard coloring (Fig.\ref{fig:sublattices}). 

\begin{figure}[h!]
\centering
\includegraphics[scale=0.5]{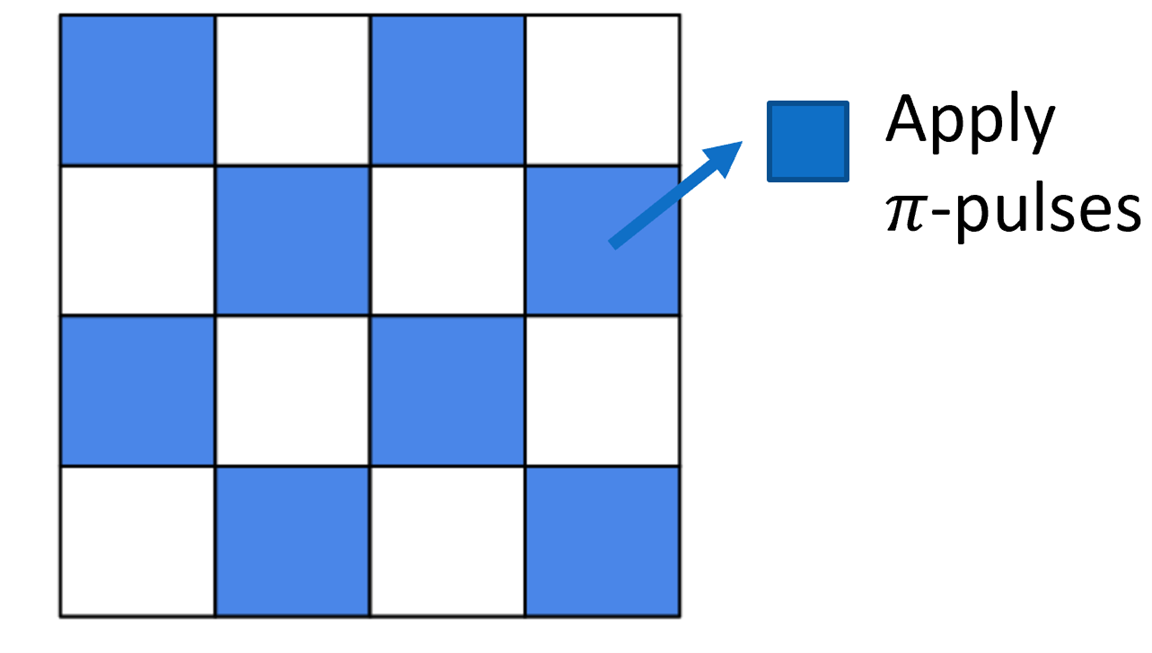}
\caption{Checkerboard partition of the 2D lattice. We apply $\pi$-pulses at the $\mathcal{K}$ sites (shown in blue) to effectively turn off the interaction in the $x$ and $y$ directions upon time-averaging.}
\label{fig:sublattices}
\end{figure}

For the set of $\mathcal{K}$ spins, we turn on a $\pi$-pulse in the middle of the quench ($k\in \mathcal{K}$), such that: 
\begin{equation}
J_{k}^m(t)=\begin{cases}
          0 \quad &\text{if} \, 0 \leq t<\frac{t_m}{2} \\
          \frac{\pi}{2\Delta t} \quad &\text{if} \, \frac{t_m}{2} \leq t<\frac{t_m}{2} + \Delta t \\
          0 \quad &\text{if} \, \frac{t_m}{2} + \Delta t \leq t<t_m \\
     \end{cases}
\end{equation}
where $\Delta t \ll t_m$ is a short duration of time. With this choice of disorder, the time evolution operator reduces to 
\begin{equation}
    \hat{\mathcal{U}} = e^{-i\hat{\mathcal{H}}_{\text{XXZ}}\Delta t}(\Pi_{k} i \hat{Z}_{k})e^{-i\sum_k B_k^m \hat{Z}_k}.
\end{equation}
Now the Pauli $Z$'s effectively flip the signs of the $\hat{S}^x\hat{S}^x$ and $\hat{S}^y\hat{S}^y$ terms in the XXZ-Hamiltonian, and upon integrating over the duration of a quench cancels out with the corresponding terms in first half of the quench. Therefore, after time evolution of a quench, the effective Hamiltonian is left with only Ising interactions, 
\begin{equation}
    \bar{\mathcal{H}}_{\text{eff}}^m = \sum_{\langle i,j \rangle} J_{zz} \hat{S}^z_i \hat{S}^z_j + \sum_i B_i^m \hat{S}^z_{i}
\end{equation}

\end{proof}

\section{MBL phase transition}
\label{app:MBL_check}

In this section, we present the details of the numerical simulation of XXZ model (Eqn.\eqref{eqn:XXZ}) and confirm the thermal to MBL phase transition. We simulate the XXZ model using exact diagonalization methods provided by the QuSpin package \cite{weinberg2017quspin,weinberg2019quspin}. Throughout the paper we use parameters $J_{xy}=J_{zz}=1$. 

\begin{figure}[h!]
\centering
\includegraphics[scale=0.25]{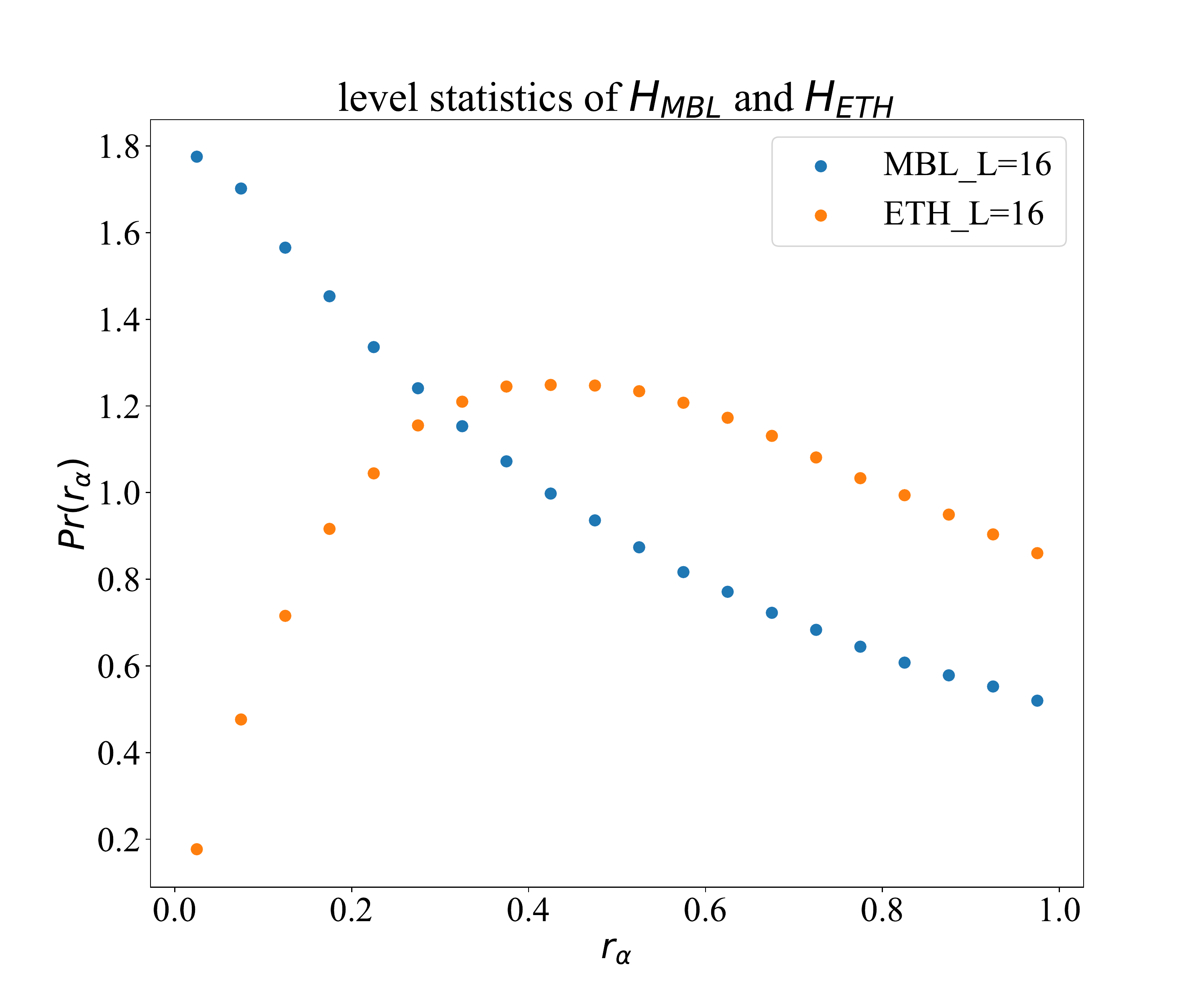}
\caption{Level statistics of $L=16$ XXZ model subject to quenches in the $z$-direction. The thermal phase (denoted as ETH) is simulated with $h_d=0.1$, with resulting $\langle r_\alpha \rangle \approx 0.391$; the MBL phase is simulated with $h_d=3.9$, with resulting $\langle r_\alpha \rangle \approx 0.529$. Results are averaged over 1000 different realizations.}
\label{fig:level_statistics_L=16}
\end{figure}

One hallmark of the MBL phase is the Poission distribution of level spacings in the eigenspectrum of the Hamiltonian. \cite{nandkishore2014many,alet2018many,Abanin_2019}. The level statistics $Pr(r_{\alpha})$ is defined as the normalized distribution of
\begin{equation}
    r_{\alpha} = \frac{\min (\Delta_{\alpha+1},\Delta_{\alpha})}{\max (\Delta_{\alpha+1},\Delta_{\alpha})},
\end{equation}
where $\Delta_{\alpha} = E_{\alpha+1}-E_{\alpha}$ are the level spacings in the eigenspectrum. In Fig.\ref{fig:level_statistics_L=16}, we show the level statistics in a simulation of $L=16$ spins described by Eqn.\eqref{eqn:H_total} subject to a single quench $M=1$, at two different disorder strengths: $h_d=0.1$ and $h_d=3.9$ (the critical disorder strength is $h_c \sim 3.5$ for $J_{zz}=J_{xy}=1$). We see that indeed the level statistics in the thermal phase obeys Wigner-Dyson statistics ($\langle r_\alpha \rangle \approx 0.391$), and in the MBL phase obeys Poisson statistics ( $\langle r_\alpha \rangle \approx 0.529$), confirming the existence of thermal-MBL phase transition. 

\begin{figure}[h!]
\centering
\includegraphics[scale=0.26]{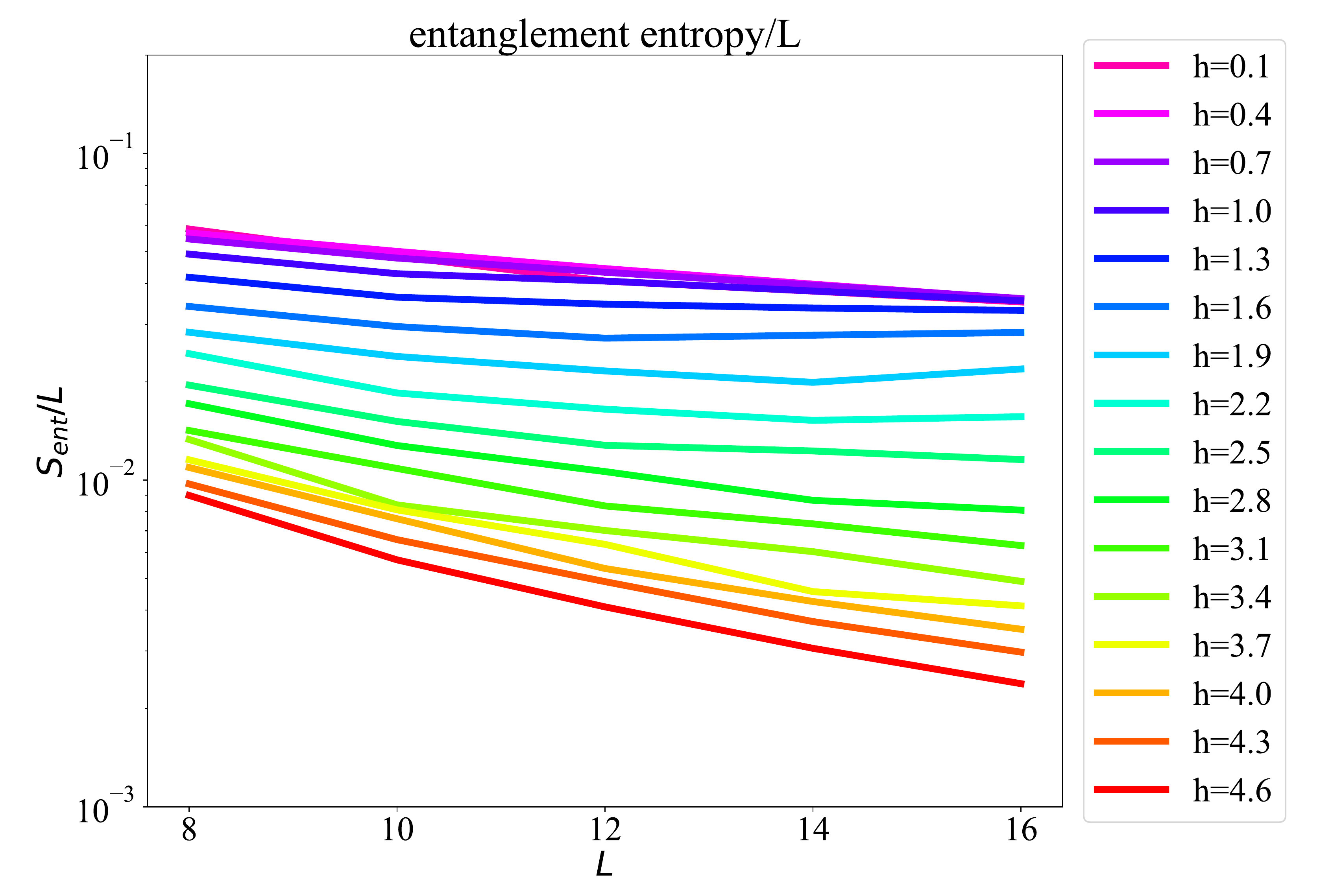}
\caption{Scaling analysis of entanglement entropy. We plot the entanglement entropy per site $S_{ent}/L$ as a function of system size $L$ for different disorder strengths $h$. Volume law scaling in the thermal phase (small $h$) leads to constant $S_{ent}/L$, while area law scaling in the MBL phase (for large $h$) leads to decreasing $S_{ent}/L$. }
\label{fig:ent_entropy_scaling}
\end{figure}

Another hallmark of MBL phase is the area law scaling of von Neumann entanglement entropy ($S_{\text{ent}}=-\rm Tr\rho \ln \rho$), compared to the volume law scaling in the thermal phase. We numerically calculate the half-system entanglement entropy in the middle of the spectrum for the Hamiltonian in Eqn.\eqref{eqn:H_total}, and perform a scaling analysis for different $L$ and different disorder strengths $h$ (see Fig.\ref{fig:ent_entropy_scaling}). Our numerical results agrees with those reported in \cite{MBL_edge}.

\section{Training MBL hidden Born machine}
\label{app:MMD}
\subsection{MMD loss}
Previously, KL-divergence has been suggested for training MBL Born machine as a generative model \cite{tangpanitanon2020expressibility}. However, KL-divergence does not capture correlations within data, and suffers from infinities outside the support of data distribution. Moreover, KL-divergence requires full knowledge of data distirbution which is often unrealistic. To remedy these situations, the Maximum Mean Discrepancy (MMD) loss has been proposed for training Born machines \cite{liu2018differentiable}. The MMD loss measures the distance between model distribution $p$ and target distribution $q$, by comparing their mean embeddings in the feature space, and one can use samples to estimate the loss. The (squared) MMD loss can be written as
\begin{align}
    \mathcal{L}_{MMD} &= \norm{\sum_x p(x)\phi(x)  - \sum_x q(x)\phi(x)}^2 \nonumber\\
    &= \mathbb{E}_{x,x'\sim p} k(x,x') + \mathbb{E}_{y,y'\sim q} k(y,y') \\
    &- 2\mathbb{E}_{x\sim p, y\sim q} k(x,y) \nonumber,
\end{align}
where we have employed the kernel trick and write $k(x,y) = \phi(x)^T \phi(x)$. In our model, we use a Gaussian mixture kernel $k(x,y) = \frac{1}{c}\sum_{i=1}^c \exp\left(-\frac{1}{2\sigma_i^2}|x-y|^2\right)$ of four channels $c=4$, with corresponding bandwidths $\sigma_i^2 = [0.1,0.25,4,10]$. The bandwidths are chosen such that our Gaussian kernels are able to capture both the local features and the global features in the target distribution.

\subsection{Parameters}
In the training of our MBL hidden Born machine, we use $N=6+2$ ($6$ visible spins and $2$ hidden spins), and $M=100$ quenches and search over $N=500$ different disorder realizations. 

Generally, we found that more hidden units lead to better learning outcome. However, for the tasks considered in this study, further increasing the number of hidden units beyond two does not result in substantial improvements in the results. Nonetheless, we expect that increasing the number of hidden units may prove beneficial for more complicated tasks, as suggested in the future directions.

\subsection{Runtime scaling}

The runtime complexity of our learning algorithm is primarily determined by its use of Monte Carlo searches, resulting in an overall complexity of $\mathcal{O}(TNM)$. Here, $T$ represents the evolution time of the system during each quench, $N$ is the task-dependent number of Monte Carlo samples, and $M$ denotes the constant number of quenches. Notably, the specific influence of $T$ depends on the simulation method employed. In the case of exact diagonalization, $T$ scales exponentially with system size, leading to substantial computational requirements for larger systems. However, it is anticipated that near-term analog or digital quantum computers will significantly enhance the efficiency of these operations, mitigating the impact of this scaling behavior.

\section{Data encoding}
\label{app:MNIST}

Here, we describe the detailed data encoding scheme and our toy dataset of MNIST digit patterns in this section. Given a reduced density matrix $\rho_{\text{vis}}$ of $L$ visible spins, we compute the distribution of finding each of the $2^L$ basis states in our computational basis, and interpret the result as pixel values. We then reshape this probability vector into an image of size $2^{L/2} \times 2^{L/2}$.

On the other hand, given an image $\vec{x}_{\mu} \in \mathbb{R}^{n\times n}$, where $n\times n$ is the number of original pixels in the image, we first downsample it to $2^{L/2} \times 2^{L/2}$ pixels, then normalized the pixel values to be within $0$ and $1$. 

Our toy dataset of MNIST digit patterns are constructed as follows: we take all the training images $\vec{x}^{\mu}$ from a digit class, downsample to $2^{L/2} \times 2^{L/2}$ pixels, and compute each pixel as the average value $\bar{x}_i = 1/P \sum_{\mu=1}^P x^{\mu}_i$ across different styles within this digit class, where $i=1,...,2^L$. We then normalized the pixels to $\bar{x}_i \to \bar{x}_i/\sum_i \bar{x}_i$ and interpret the result as $q_{\text{data}}$. We take caution that this is different from learning the MNIST distribution in generative models. The latter refers to learning the joint probability distribution over all pixels in the image, and our toy data set corresponds to taking the mean-field limit of this joint probability distribution, which ignores the complicated correlations among pixels. This is akin to learning a single pattern (the averaged MNIST images shown in Fig.\ref{fig:MNIST_0_9}), and the reason for taking the average pixel value is such that we will be able to perform pattern recognition with imperfect initial states.

\end{document}